\documentclass[english,aps,manuscript,amsmath,amssymb]{revtex4}
\usepackage[latin9]{inputenc}
\usepackage{amsmath}
\usepackage{amssymb}
\usepackage{graphicx}
\usepackage{esint}

\makeatletter
\@ifundefined{textcolor}{}
{%
 \definecolor{BLACK}{gray}{0}
 \definecolor{WHITE}{gray}{1}
 \definecolor{RED}{rgb}{1,0,0}
 \definecolor{GREEN}{rgb}{0,1,0}
 \definecolor{BLUE}{rgb}{0,0,1}
 \definecolor{CYAN}{cmyk}{1,0,0,0}
 \definecolor{MAGENTA}{cmyk}{0,1,0,0}
 \definecolor{YELLOW}{cmyk}{0,0,1,0}
}

\makeatother

\usepackage{babel}
\begin{document}

\title{Nonequilibrium Glass Transition in Mixtures of Active-Passive Particles}

\author{Huai Ding, Mengkai Feng, Huijun Jiang, Zhonghuai Hou}

\thanks{Corresponding Author: hzhlj@ustc.edu.cn}

\affiliation{Department of Chemical Physics \& Hefei National Laboratory for Physical
Sciences at Microscales, iCheM, University of Science and Technology
of China, Hefei, Anhui 230026, China}
\begin{abstract}
We develop a mode coupling theory(MCT) to study the nonequilibrium
glass transition behavior of a mono-disperse mixture of active-passive
hard-sphere particles. The MCT equations clearly demonstrate that
the glass transition is shifted to higher values of total volume fraction
when doping a passive system with active particles. Interestingly,
we find that the glass transition point may show a non-monotonic dependence
on the effective diffusivity of the active component, indicating a
nontrivial type of activity induced reentrance behavior. Analysis
based on the nonergodic parameters suggest that the glassy state at
small activity is due to the caging effect, while that at high activity
could result from activity induced dynamic clustering.
\end{abstract}
\maketitle

\section{introduction}

The collective behavior of systems containing active particles have
gained extensive attention in recent years due to its great importance
both from a fundamental physics perspective and for understanding
many biological systems\cite{2012_PhysRep_CollectiveMotion,2013_RMP_Hydro}.
A wealth of new nonequilibrium phenomena have been reported, such
as active swarming, large scale vortex formation\cite{2010_Nature_SoftFilament,2012_Nature_HardFilament},
phase separation\cite{2012_PNAS_Phase_separation,2012_PRL_AthermalPhaseSeparation,2013_PRL_PhaseSeparation,2014_NatComm_Phi4Theory,2015_PRL_AP_Mixture},
etc, both experimentally and theoretically. Recently, a new trend
in this field has been the dynamics of dense assemblies of self-propelled
particles around glass transition. Experiments on crowded systems
of active colloids and active cells show dynamic features such as
jamming and dynamic arrest that are very similar to those observed
in glassy materials \cite{2010_PRL_Active_Sedimentation,2011_PNAS_Exp_Cells}.
Computer simulations demonstrated that nonequilibrium glass transition
or dynamic arrest behavior does occur in a dense suspension of self-propelled
hard spheres, where the critical density for glass transition shifts
to larger value with increasing activity\cite{2013_NC_NiRan,2014_PRL_GT_ActiveHardDisk}
. Starting from a generalized Langevin equation with colored non-thermal
noise, L.Breather and J. Curtain theoretically predicted that dynamic
arrest can occur in systems that are far from equilibrium, showing
that non-equilibrium glass transition moves to lower temperature with
increasing activity and to higher temperature with increasing dissipation
in spin glasses\cite{2013_NatPhys_Kurchan}. Mode coupling theory
(MCT) were also proposed recently to study the glassy dynamics of
driven granular fluids\cite{2010_PRL_GT_DrivenGranularFluids} and
active colloidal suspensions\cite{2014_arXiv_MCT_ActiveGlass}.

While most of the studies so far have only considered single-component
active particles, very recently, mixture systems of active-passive
particles began to draw new attentions. Interesting experiments\cite{2012_PNAS_Phase_separation,2013_SCIENCE_Living_Crystals}
reported that the addition of active particles in a system can dramatically
alter its phase behavior. Molecular dynamics simulation showed that
introduction of activity to a passive system may not only hamper phase
separation, but can enhance it as well, based on the coordination
among the active particles \cite{2014_PRL_AP_Mixture}. It was also
demonstrated that activity can induce phase separation and direct
self-assembly in active-passive mixtures\cite{2015_PRL_AP_Mixture}.
In particular, Brownian simulations showed that one may crystallize
hard-sphere glass by doping with active particles\cite{2014_SoftMatt_NiRan},
which proposes a very interesting question about the glass transition
behavior of active-passive mixtures. However, an unified microscopic
theoretical framework to describe this important issue is still lacking.

In the present paper, we develop a general MCT framework to study
the glass transition behavior of active-passive mixtures. Our starting
point is the Smoluchowski equation for $N$-particle probability density
function, wherein the particle activity is realized via an effective
diffusivity that is larger than that of passive particles. Such a
treatment allows one to apply the Mori-Zwanzig projection operator
formalism such that a set of closed equations regarding the time evolutions
of the density correlators can be obtained. In particular, we apply
this approach to a binary mixture of mono-disperse active-passive
hard sphere particles, with particular attention paid on how the particle
activity and the number fraction of active particles would influence
the glass transition behaviors. While doping with active particles
can shift the transition to higher total volume fraction as expected,
we find that increasing particle activity may lead to an nontrivial
type of reentrance glass transition behavior.

\section{THEORY}

For generality, we consider an $m$-component mixture of $N$ spherical
colloidal particles, being active or not, dispersed in a simple fluid
with temperature $T$ and volume $V$. For passive particles, the
over-damped dynamics can be described by Langevin equations involving
the time evolution of the position vector $\mathbf{r}_{l}^{\mu}$
for an $\mu$-type particle labeled $l$. For active particles, one
generally needs to consider a further orientation variable $\vartheta_{l}^{\mu}$
to account for the tumbling or rotational diffusion\cite{2012_RPP_Diffusive_transport,2012_EPL_Active_2D_traps}.
Very recently, it was demonstrated that the dynamics of an active
particle undergoing self-propulsion and rotational diffusion can be
well-approximated by a random-walk in the long time limit with an
effective diffusivity\cite{2012_RPP_Diffusive_transport,2013_EPL_effective_diffusion}.
These observations facilitate us to propose a minimal model for active-passive
mixtures, which starts from the Smoluchowski equation for the probability
density $P\left(\mathbf{r}^{N},t\right)$ of the particle configuration
$\mathbf{r}^{N}=\left\{ r_{l}^{\mu}\right\} _{l=1,...,N_{\mu};\mu=1,...,m}$,

\begin{equation}
\partial P\left(\mathbf{r}^{N},t\right)/\partial t=\hat{\mbox{\ensuremath{\Omega}}}P\left(\mathbf{r}^{N},t\right)\label{eq:SE}
\end{equation}
where $\hat{\Omega}$ is the Smoluchowski operator
\begin{equation}
\hat{\Omega}=\sum_{\mu=1}^{m}\sum_{j=1}^{N_{\mu}}D_{0}^{\mu}\nabla_{j}^{\mu}\cdot\left(\lambda^{\mu}\nabla_{j}^{\mu}-\beta\mathbf{F}_{j}^{\mu}\right)\label{eq:SO}
\end{equation}
Here $D_{0}^{\mu}$ denotes the bare diffusivity for $\mu$-type particles
with total number $N_{\mu}$ and $\beta=1/k_{B}T$ with $k_{B}$ the
Boltzmann constant. $\mathbf{F}_{j}^{\mu}=-\nabla_{j}^{\mu}U\left(\mathbf{r}^{N}\right)$
is the direct force acting on the particle $j\in\mu$ due to the total
potential energy $U\left(\mathbf{r}^{N}\right)$ of the colloidal
particles, and $\nabla_{j}^{\mu}$ is the gradient operator with respect
to $\mathbf{r}_{j}^{\mu}$. A key factor here is the parameter $\lambda^{\mu}$
characterizing the particle activity of $\mu$-species: $\lambda^{\mu}>1$
for active particles while $\lambda^{\mu}=1$ for passive ones. For
an equilibrium distribution $P_{e}\left(\mathbf{r}^{N},t\right)\propto e^{-\beta U}$,
$\nabla_{j}^{\mu}P_{e}=+\left(\beta\mathbf{F}_{j}^{\mu}\right)P_{e}$
such that the random force balances the potential force for passive
particles, but not for active ones.

To probe the collective dynamics of the system, one generally considers
the density correlator
\begin{equation}
\Phi_{\mu\nu}\left(\mathbf{q},t\right)=\left\langle \left(e^{\hat{\Omega}^{\dagger}t}\rho_{\mathbf{q}}^{\mu}\right)\rho_{\mathbf{-q}}^{\mu}\right\rangle \label{eq:Density-Correlator}
\end{equation}
 where $\rho_{\mathbf{q}}^{\mu}=\frac{1}{\sqrt{N_{\mu}}}\sum_{j=1}^{N_{\mu}}e^{-i\mathbf{q\cdot r_{j}^{\mu}}}$
is the Fourier transform of the density $\rho^{\mu}(\mathbf{r},t)=\frac{1}{\sqrt{N_{\mu}}}\sum_{j=1}^{N_{\mu}}\delta\left(\mathbf{r}-\mathbf{r}_{j}^{\mu}\right)$,
and $\left\langle \cdot\right\rangle $ denotes an equilibrium average.
$\hat{\Omega}^{\dagger}$ is the adjoint or backward Smoluchowski
operator given by\cite{1996_PhysRep_Nagele,1999_JCP_Nagele,2014_arXiv_MCT_ActiveGlass}

\begin{equation}
\hat{\Omega}^{\dagger}=\sum_{\mu=1}^{m}D_{0}^{\mu}\sum_{j=1}^{N_{\mu}}\left(\lambda^{\mu}\nabla_{j}^{\mu}+\beta\mathbf{F}_{j}^{\mu}\right)\nabla_{j}^{\mu}\label{eq:Omega_Dagger}
\end{equation}
where for arbitrary functions $f$ and $g$ of $\mathbf{r}^{N}$ we
have $\int d\mathbf{r}^{N}f\left(\hat{\Omega}g\right)=\int d\mathbf{r}^{N}\left(\hat{\Omega}^{\dagger}f\right)g$.
The use of adjoint Smoluchowski operator facilitates us to apply Mori-Zwanzig
approach and mode-coupling methods\cite{1999_JCP_Nagele} to obtain
an approximate dynamic equation for the density correlators (\ref{eq:Density-Correlator})
which reads in the matrix form as (See the Supplemental Information)

\begin{multline}
\frac{\partial}{\partial t}\mathbb{\mathbf{\Phi}}\left(q,t\right)+\mathbf{W}\left(q\right)\cdot\mathbf{S}^{-1}\left(q\right)\cdot\mathbf{\Phi}\left(q,t\right)\\
+\int_{0}^{t}du\mathbf{M}\left(q,t-u\right)\cdot\mathbf{W}^{-1}\left(q\right)\cdot\frac{\partial}{\partial u}\mathbf{\Phi}\left(q,u\right)=0\label{eq:MCT_MatrixForm}
\end{multline}
Herein, $\mathbf{S}^{-1}\left(q\right)$ is the inverse of the partial
structure factor matrix $\mathbf{S}\left(q\right)=\mathbf{\Phi}\left(q,t=0\right)$,
$\mathbf{W}\left(q\right)$ is a frequency matrix given by
\begin{equation}
W_{\mu\nu}\left(q\right)=q^{2}D_{0}^{\mu}\left[\left(\lambda^{\mu}-1\right)S_{\mu\nu}\left(q\right)+\delta_{\mu\nu}\right].\label{eq:Fre_Matrix}
\end{equation}
$\mathbf{M}\left(q,t\right)$ denotes the matrix of so-called $irreducible$
memory function\cite{1987_Cichocki_irreducible,1994_Kawasaki_irreducible,1999_JCP_Nagele}
with elements given by

\begin{eqnarray}
M_{\mu\nu}\left(q,t\right) & = & \frac{1}{2}\sum_{\mathbf{k}}\sum_{\delta\gamma\delta'\gamma'=1}^{m}V_{\mu;\delta\gamma}\left(\mathbf{q},\mathbf{k}\right)V_{\nu;\delta'\gamma'}\left(\mathbf{q},\mathbf{k}\right)\nonumber \\
 &  & \text{\ensuremath{\times}}\Phi_{\delta\delta'}\left(k,t\right)\Phi_{\gamma\gamma'}\left(\left|\mathbf{q}-\mathbf{k}\right|,t\right)\label{eq:Memory_Kernel}
\end{eqnarray}
with vortex functions

\begin{eqnarray}
V_{\mu;\delta\gamma}\left(\mathbf{q},\mathbf{k}\right) & = & \frac{\rho D_{0}^{\mu}}{\sqrt{N_{\mu}}}[\left(\mathbf{q}\cdot\mathbf{k}\right)\delta_{\mu\gamma}C_{\mu\delta}\left(\mathbf{k}\right)\nonumber \\
 &  & +\mathbf{q}\cdot\left(\mathbf{q-k}\right)\delta_{\mu\delta}C_{\mu\gamma}\left(\mathbf{\left|q-k\right|}\right)]\label{eq:Vortex}
\end{eqnarray}
where $\rho=N/V$ is the total number density and $C_{\mu\nu}\left(k\right)=\sqrt{x_{\mu}x_{\nu}}c_{\mu\nu}\left(k\right)$
in which $c_{\mu\nu}\left(k\right)$ are direct correlation functions.
$C_{\mu\nu}\left(k\right)$ is related to the static structure factor
via $\rho C_{\mu\nu}=\delta_{\mu\nu}-\left(\mathbf{S}^{-1}\right)_{\mu\nu}$
.

The MCT equations (\ref{eq:MCT_MatrixForm}) to (\ref{eq:Vortex})
for general active-passive mixtures constitute the first part of central
results of the present paper. In general, one may employ them to study
the glass transition or dynamic arrest behavior of any multi-component
systems. In the present work, we will mainly focus on a two-component
hard-sphere system, one is active labeled by '$a$' and the other
is passive labeled by '\textbf{$p$}' . For simplicity, we consider
that the diameters $d$ of both types of particles are the same. The
total volume fraction is given by $\varphi=\rho\pi d^{3}/6$, with
$d$ set to be 1. The number fraction and the activity parameter of
the active component are given by $x_{a}$ and $\lambda_{a}$ respectively,
which are chosen as the main control parameters in the present study.
Eqs.(\ref{eq:MCT_MatrixForm}) are numerically solved with Percus-Yevick
static structure factors $S_{\mu\nu}\left(k\right)$ $(\mu,\nu=a\mbox{ or }p)$
as input and setting $D_{0}^{a}=D_{0}^{p}=1$. For a pure passive
system with $x_{a}=0$, the equations predict a glass transition (GT)
at volume fraction $\varphi_{c}^{0}=0.515$.

\section{results and discussion}

\begin{figure}
\centering{}\includegraphics[width=1\columnwidth]{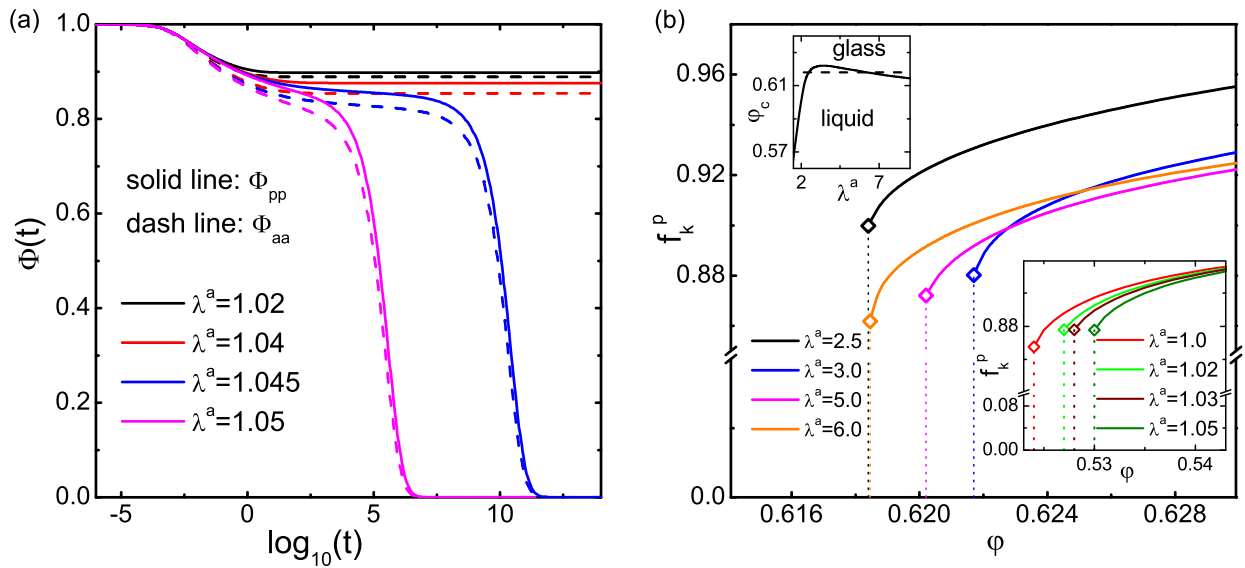}\protect\caption{(Color online) (a) Dynamic scattering functions $\Phi_{pp}\left(k,t\right)$
for passive particles (solid lines) and $\Phi_{aa}\left(k,t\right)$
for active particles (dashed lines) evaluated at $k=k_{m}$ for different
values of the activity parameter $\lambda^{a}$. The total volume
fraction is $\varphi=0.524$ and the number fraction of active particles
is $x_{a}=0.5$. (b) Dependences of the non-ergodic parameters $f_{k}^{p}$
evaluated at $k=k_{m}$ on the total volume fraction $\varphi$ for
different values of $\lambda^{a}$ and $x_{a}=0.5$. The right inset
shows $f_{k}^{p}$ for several values of $\lambda^{a}$ close to 1.0
for clarity. The left inset plots the dependence of $\varphi_{c}$
on $\lambda^{a}$ where a turnover phenomenon is apparent, wherein
the dashed line refers to $\varphi=0.618$ where the system shows
reentrance behavior. See the text.}
\label{Fig_Phi_t_Lambda}
\end{figure}

First of all, we study how the particle activity influences the GT.
To this end, we fix $x_{a}=0.5$ and set $\varphi=0.524$ which is
above the GT point $\varphi_{c}^{0}$ for pure passive system. In
Fig.\ref{Fig_Phi_t_Lambda}(a), we show the time evolution of the
dynamic scattering functions $\Phi_{aa}\left(k,t\right)$ and $\Phi_{pp}\left(k,t\right)$
evaluated at $k=k_{m}$, where $k_{m}\simeq7.23$ is the location
of the first peak in $S_{k}$, with varying particle activity $\lambda^{a}$
in a relatively small range. When $\lambda^{a}$ is just slightly
than one, e.g., $\lambda^{a}=1.02$, both correlators do not decay
in the long time and the system remain in the glassy state. The nonegodicity
parameter $f_{k}^{a}=\Phi_{aa}\left(k,t\rightarrow\infty\right)$
for the active particle (dashed line) is slightly slower than $f_{k}^{p}=\Phi_{pp}\left(k,t\rightarrow\infty\right)$
for the passive particle, indicating that the glassy part formed by
the active component is softer than the passive part. With increasing
$\lambda^{a}$, the plateau heights of both correlators decrease.
Above some threshold value of $\lambda^{a}$(about 1.044 here), the
correlators finally decay to zero and the system becomes fluid, wherein
the active component relaxes faster than the passive one.

Above results clearly demonstrate that increasing activity of the
active component can shift GT to higher volume fractions. In Fig.\ref{Fig_Phi_t_Lambda}(b),
we plot the nonergodic parameters $f_{k}^{p}$ evaluated at $k=k_{m}$
as functions of the total fraction $\varphi$ for several different
values of $\lambda^{a}$. As it should, $f_{k}^{p}$ takes a finite
value at a certain discontinuous glass transition point $\varphi_{c}$
before which $f_{k}^{p}=0$. Clearly, the value of $\varphi_{c}$
increases fast with $\lambda^{a}$ at first, e.g., $\varphi_{c}\simeq0.515$
for $\lambda^{a}=1.0$ and $\varphi_{c}\simeq0.622$ for $\lambda^{a}=3.0$.
Strikingly, however, with further increasing of $\lambda^{a}$ to
larger values like 5.0, $\varphi_{c}$ decreases again to a smaller
value around 0.620. In the left inset of Fig.\ref{Fig_Phi_t_Lambda}(b),
the dependence of $\varphi_{c}$ on $\lambda^{a}$ is shown, where
the turnover phenomenon is apparent.

\begin{figure}
\centering{}\includegraphics[width=1\columnwidth]{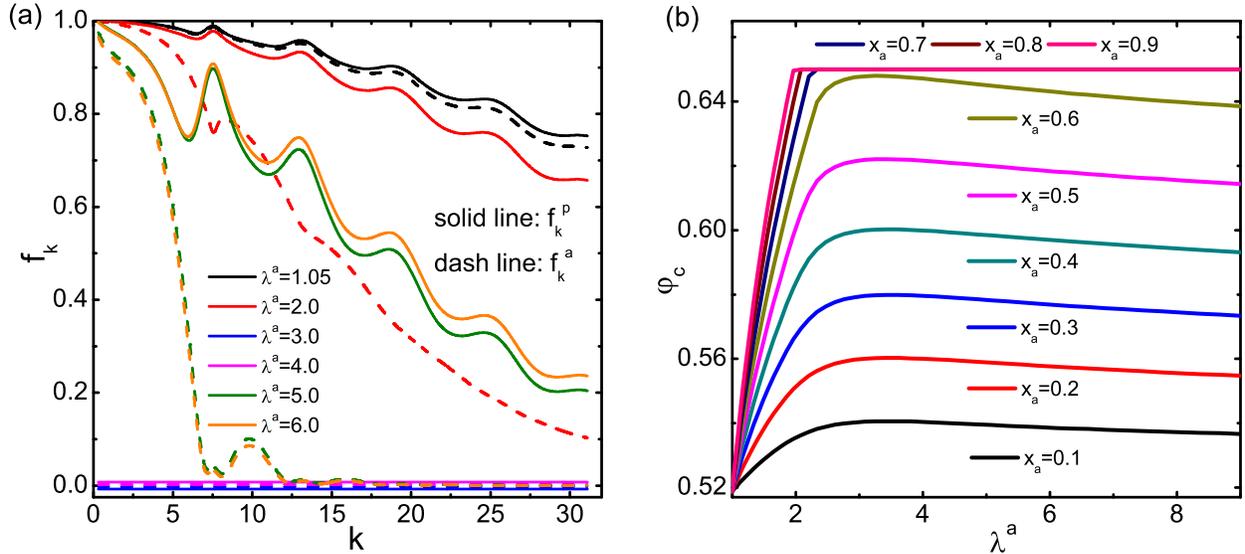}\protect\caption{(Color Online) (a)Non-ergodic parameters $f_{k}^{p}$ (solid lines)
and $f_{k}^{a}$ (dashed lines) in the whole $k$ range for different
values of $\lambda^{a}$ at fixed total volume fraction $\varphi=0.618$
and $x_{a}=0.5$ where reentrance behavior can be observed. For relatively
large $\lambda^{a}=5.0$ and $6.0$, $f_{k}^{a}$ is nearly zero for
large $k$ and apparently nonzero for small $k$. (b) Dependence of
the glass transition point $\varphi_{c}$ on particle activity $\lambda^{a}$
for different $x_{a}$. Reentrance behavior disappears if $x_{a}$
is too large. }
\label{Fig_fk_All}
\end{figure}

The above findings indicate an interesting type of reentrance behavior
in active-passive mixtures for a fixed total volume fraction $\varphi$.
For instance, for $\varphi=0.618$ as indicated by the dashed line
in the left inset of Fig.\ref{Fig_Phi_t_Lambda}(b), the system changes
first from glass to liquid and then to glass again as $\lambda^{a}$
increases. While the first transition from glass to liquid is as expected
because activity can push the glass transition to higher volume fraction,
the second one from liquid to glass again is rather counterintuitive.
To get more insight, we have plot the nonergodic parameters $f_{k}^{a}$
and $f_{k}^{p}$ for all wave vectors in Fig.\ref{Fig_fk_All}(a)
for $\varphi=0.618$ and $\lambda^{a}$ from 1.05 to 6.0. For $\lambda^{a}=1.05$
where the system is in glassy state, we see that the profiles of$f_{k}^{a}$
and $f_{k}^{p}$ are nearly the same, with only slight differences
in their values. For $\lambda^{a}=2.0,$ the system is still glass,
with the active part much softer than the passive part. For $\lambda^{a}=3.0$
and 4.0, the system is in the liquid phase because all the $f_{k}$s
are now zero. Nevertheless, for further larger values of $\lambda^{a}$,
say 5.0 and 6.0, the system becomes glassy again since both $f_{k}^{a}$
and $f_{k}^{p}$ have nonzero values in some $k$ range. Interestingly,
we find that the profiles of $f_{k}^{a}$ and $f_{k}^{p}$ are quite
different in this latter case. $f_{k}^{p}$ remains a relatively large
value and contains many peaks indicating the passive component is
frozen at all length scales but with a certain type of structure.
However, $f_{p}^{a}$ is only apparently nonzero for small $k$ and
is almost zero for $k$ larger than some threshold value, suggesting
that the active part is still liquid-like in short length scales but
being frozen in large scales.

We note here that a few interesting reentrance behaviors regarding
glass transition have been reported in the literatures. For instance,
for passive hard-spheres with short-range attractions, it was found
that increasing the attraction can melt glass, however, further increasing
attraction can lead to glass again\cite{2002_Science_MultipleGlass}.
The reentrance was due to the existence of two qualitatively different
glassy states: one with structural arrest due to caging and the other
with arrest due to bonding. For passive hard sphere mixtures with
very disparate sizes, reentrance behavior\cite{2003_PRE_reentrance_behavior}
as well as multiple glasses\cite{2011_EPL_Mixture} have also been
reported. Reentrance glass transition has also been found for fluids
in porous media, where for large volume fraction of immobile matrix
particles, increasing or decreasing the number fraction of fluid particles
may both lead to glassy states\cite{2005_PRL_porous,2007_PRE_porous}.
Multiple reentrant glass transitions were also found in a confined
system of hard spheres between two parallel walls\cite{2014_NatComm_ReGT}.
Here we report a new type of reentrance glass transition behavior
induced by particle activity. As demonstrated in Fig.\ref{Fig_fk_All},
the reentrance here may be also due to existence of two types of nonequilibrium
glassy states. For small activity, the dynamic arrest is due to the
caging effect of both passive and active particles. Increasing activity
acts as 'noise' which can destroy the cages and melt the glass. If
activity is too large, however, local phase separation\cite{2012_PNAS_Phase_separation,2012_PRL_AthermalPhaseSeparation,2013_PRL_PhaseSeparation,2014_NatComm_Phi4Theory,2015_PRL_AP_Mixture}
of the active particles could happen such that active particles can
form local clusters, which may also leads to dynamic arrest. In this
case, the non-clustered active particles can still move like liquids
within small length scales, such that $f_{k}^{a}$ is nearly zero
for large $k$. It is interesting to check such a scenario by large
scale computer simulations, which definitely deserves a separate future
work. We would like to mention here that in a recent simulation work
of active-passive mixtures\cite{2015_PRL_AP_Mixture}, the authors
demonstrated that the active component triggers phase separation into
a dense and a dilute phase, where active-passive segregation were
further observed in the dense phase with rafts of passive particles
in a sea of active particles. We suggest that the reentrant glassy
state for large activity we found in the present work may look like
this type of active-passive segregation in dense phase.

We now consider how the above results depend on the number fraction
$x_{a}$ of the active component. In Fig.\ref{Fig_fk_All}(b), the
dependences of $\varphi_{c}$ on $\lambda^{a}$ for different values
of $x_{a}$ are shown. Clearly, the reentrance behavior can occur
in a large range of values of $x_{a}$. But for too large $x_{a}$,
$\varphi_{c}$ will become saturated to a value about 0.65, which
is consistent with the result for a pure active system\cite{2014_arXiv_MCT_ActiveGlass}.
Thus one conclude that the interesting reentrance behavior reported
above is a specific feature of active-passive mixture system and cannot
occur for a pure active system. The data also indicates that $\varphi_{c}$
increases monotonically with $x_{a}$ for a fixed $\lambda^{a}$,
which is in consistent with the simulation work of R. Ni $et\, al.$
that doping active particles can help crystallizing hard spheres by
melting glass\cite{2014_SoftMatt_NiRan}. Note that the reentrance
behavior does not exist with variation of $x_{a}$.

\section{conclusion}

In conclusion, we have developed a mode-coupling theory starting from
Smoluchowski equations to study the nonequilibrium glassy dynamics
in mixtures of active-passive particles. This microscopic theory makes
it convenient to study the relaxation of the density correlators and
make predictions about the liquid-glass transition boundaries. In
particular, we have applied our theory to investigate the glass dynamics
of a binary mixture of mono-disperse active-passive hard spheres.
The theory clearly demonstrates that doping with active particles
will push the critical volume fraction $\varphi_{c}$ for glass transition
to higher values. In addition, we find an interesting type of reentrance
behavior induced by particle activity. For a certain given total volume
fraction where the system is in glassy state without activity, increasing
the activity level will first melt the glass, while further increasing
activity may lead to a glassy state again. The $k$-dependent nonergodic
parameters for the active and passive components share similar profiles
in the former glass state, while they are quite different in the latter
one, wherein the active particles are frozen in large scales but remain
fluid-like in small ones. Such a reentrance behavior is found to be
a specific feature of a mixture system, while it is absent in a pure
active system. We believe that our present work can offer more perspectives
in the study of collective dynamics of active-passive mixtures, as
well as the frontier topic regarding nonequilirbium glass transition.
\begin{acknowledgments}
This work is supported by National Basic Research Program of China
(2013CB834606), by National Science Foundation of China (21125313,
21473165, 21403204), and by the Fundamental Research Funds for the
Central Universities (WK2060030018, 2340000034).
\end{acknowledgments}


\end{document}